\documentclass[twoside]{cwireport}

\usepackage{graphics}
\usepackage[english]{babel}
\usepackage{lhs2TeXPreamble}
\usepackage{url}

\newcommand{\JOOS}{JOOS}
\newcommand{\mysection}[1]{\bigskip\section{#1}\smallskip}
\newcommand{\mysubsection}[1]{\medskip\subsection{#1}\smallskip}
\newcommand{\myparagraph}[1]{\medskip\paragraph{#1}\smallskip}

\begin{document}

\title{Towards Generic Refactoring\\
{\normalsize \today}}

\author{Ralf L{\"a}mmel}

\maketitle

%%%%%%%%%%%%%%%%%%%%%%%%%%%%%%%%%%%%%%%%%%%%%%%%%%%%%%%%%%%%%%%%%%%%%%%%%%%%%%

\begin{abstract}
We study program refactoring while considering the language or even
the programming paradigm as a parameter. We use typed functional
programs, namely Haskell programs, as the specification medium for a
corresponding refactoring framework.  In order to detach ourselves
from language syntax, our specifications adhere to the following
style. (I) As for primitive algorithms for program analysis and
transformation, we employ generic function combinators supporting
generic traversal and polymorphic functions refined by ad-hoc
cases. (II) As for the language abstractions involved in refactorings,
we design a dedicated multi-parameter class. This class can be
instantiated for abstractions as present in various languages, e.g.,
Java, Prolog or Haskell.

\bigskip 

\ACCsyst{D.1.1, D.1.2, D.2.1, D.2.3, D.2.13, D.3.1, I.1.1, I.1.2, I.1.3}

\keywords{
refactoring,
program transformation,
reuse,
generic programming,
frameworks,
functional programming}

\end{abstract}

%%%%%%%%%%%%%%%%%%%%%%%%%%%%%%%%%%%%%%%%%%%%%%%%%%%%%%%%%%%%%%%%%%%%%%%%%%%%%%

\mysection{Introduction}

\myparagraph{Refactoring}

The very term refactoring has recently been pushed a lot in the
context of object-oriented programming, but the related idea of
semantics-preserving program transformation is as old as high-level
programming. A program refactoring is typically meant to improve the
internal structure of a program~\cite{Fowler99}, be it to make the
program more comprehensible, to enable its reuse, or to prepare a
subsequent adaption. In a broader sense, one might also include
program transformation in the sense of refinement or optimization. Let
us consider a standard refactoring, namely the \emph{extraction} of an
abstraction from a given program. Extraction (say, folding) introduces
a name for a previously anonymous piece of code.  Obviously, the
established abstraction creates potential for reuse. Also, the
extracted functionality is maybe more concisely documented by the
abstraction, or more accessible for a subsequent adaptation. Depending
on the language which we want to deal with, different kinds of code
fragments and abstractions are relevant. Here is a list of some
classes of languages, corresponding syntactical domains involved in
extraction, and references to previous work on program transformation
with relevance for refactoring:

\noindent
\begin{center}
\begin{tabular}{l|l|l|l}
\emph{Class of languages} & 
\emph{Focused fragment} & 
\emph{Extracted abstraction} & 
\emph{References}\\ \hline
XML/DTD & content particle & element type & \cite{LL01}\\
Logic programming & literal & predicate & \cite{PS90,PP96}\\
Preprocessing & code fragment & macro & \cite{Favre96,KR01}\\
Functional programming & expression & function & \cite{PP96,Bellegarde95,Laemmel00-SFP99}\\
OO programming & statement & method & \cite{OpdykePhD,Fowler99,LV02-PADL}\\
Syntax definition & EBNF phrase & nonterminal &
\cite{Pepper99,Laemmel01-FME}
\end{tabular}
\end{center}

\myparagraph{Genericity}

One might wonder what the commonalities of program refactorings (such
as extraction) are if we attempt to consider the language or even the
programming paradigm as a parameter. To address these problems, we
develop a language-independent refactoring framework. A
language-independent formulation of refactoring has not been suggested
or attempted before, but it turns out to be informative and useful. As
a first indication, the commonalities, which we are able to capture,
are of the following kind:
\begin{itemize}
\item There are general notions of focus, scope, and abstraction.
\item One can navigate through programs, e.g., nested lists of
abstractions.
\item There is an interface for name analyses.
\item A refactoring can be described
by a number of steps of the following kind:
\begin{itemize}
\item Identification of fragments of a certain type and location;
\item Destruction, analysis, and construction;
\item Checking for pre- and postconditions;
\item Placing, removing or replacing a focus.
\end{itemize}
\item There are parameters for language-specific ingredients.
\end{itemize}
The refactoring framework is specified in Haskell~98 \cite{Haskell98}
with one common extension which is used for convenience, namely
functional dependencies~\cite{Jones01}. We rely on the
\emph{Strafunski} style of generic functional programming~\cite{LV02-PADL}
(joint work of the author with Joost Visser; see
\url{http://www.cs.vu.nl/Strafunski/}). This approach is based on
generic function combinators including combinators for generic
traversal and update of polymorphic functions by ad-hoc cases. The
interested reader is referred to \cite{Laemmel02-RANK2} for the
foundation of generic programming with \emph{Strafunski}-like function
combinators.

\begin{figure}[ht!]
\begin{center}
\resizebox{.9\textwidth}{.35\textheight}{\includegraphics{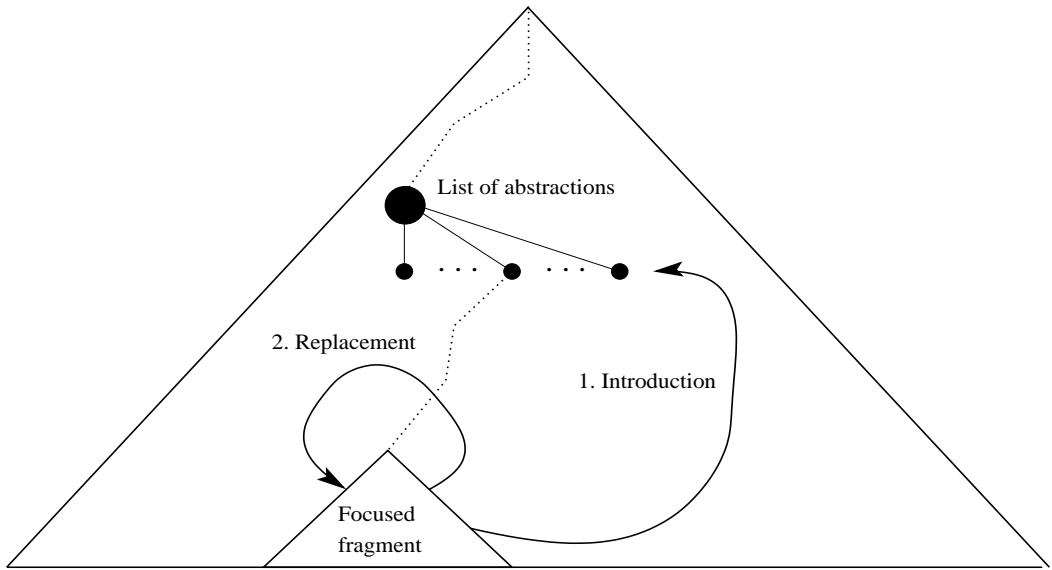}}
\end{center}
\caption{Illustration of extraction}
\label{F:extract.fig}
\end{figure}

\myparagraph{Running example}

In Figure~\ref{F:extract.fig}, we illustrate some abstract properties
of extraction. The figure shows the skeleton of a syntax tree. We
assume that there is a focused code fragment which is meant to
constitute the body of a new abstraction. The first major step of
extraction is to construct an abstraction from the focused piece of
code, and to add it to the relevant scope. Adding the new abstraction
leads to an intermediate result of extraction. We use the term
\emph{introduction} to denote the process of adding an abstraction. In
fact, introduction is another refactoring. We assume that abstractions
are hosted in possibly nested lists of abstractions. This clearly
implies that we need to be prepared for nested scopes.  The proper
scope for the new abstraction is the next list of abstractions
\emph{above} the focus. The introduction refactoring has to be
restricted not to add an abstraction which interferes with existing
ones. Once, the abstraction has been added, the focus can be replaced
by an application of the abstraction.  We have to take free names in
the focused fragment into account. These names constitute the formal
parameters of the new abstraction, and the actual parameters of the
application for focus replacement. At certain points along a
refactoring, we might have to check specific properties of some
fragments.

To give a language-specific example of extraction, consider the
extraction of a Java method. In this case, the focused piece of code
is a statement. One has to check that the potential compound statement
does not contain a \texttt{return} statement since the \texttt{return}
statement will lead to a different control-flow once placed in another
method. One also has to check that there are no assignments to
non-local variables since these side effects would not be
propagated. The focused statement constitutes the body of the
extracted method. The arguments of the emerging method are retrieved
by a free variable analysis on the focused piece of code. The focused
statement will be finally replaced by a method invocation.

\myparagraph{Schedule}

In Section~\ref{S:sf}, the Strafunski style of generic functional
programming is briefly recalled. This style is crucial for our
specification of language-independent refactorings. In
Section~\ref{S:framework}, the framework for refactoring is worked
out. We aim at a concise specification where the framework is equipped
with a number of hot spots using different parameterization
techniques. In Section~\ref{S:joos}, an instantiation of the framework
for a Java-like language is worked out.  In Section~\ref{S:concl}, the
paper is concluded.

\myparagraph{Acknowledgement}

I am very grateful for the collaboration with Joost Visser in the
\emph{Strafunski} project. The author was supported in part by the
Dutch research organisation NWO in the project 612.014.006
("\emph{Generation of program transformation systems}"). The author
gave presentations on generic refactoring at the CWI meeting on
language design assistants, action semantics, modular language
definitions in Amsterdam, The Netherlands, August 24, 2001, and at the
56th IFIP WG 2.1 meeting ("\emph{Algorithmic languages and calculi}")
on Ameland, The Netherlands, September 10-14, 2001.

%%%%%%%%%%%%%%%%%%%%%%%%%%%%%%%%%%%%%%%%%%%%%%%%%%%%%%%%%%%%%%%%%%%%%%%%%%%%%%

\mysection{Generic functional programming}
\label{S:sf}

Strafunski-like generic functions \cite{LV02-PADL,Laemmel02-RANK2}
enable a combinator-based approach to typeful generic functional
programming which is particularly suited for generic program schemes
dealing with term traversal. In this section, we briefly recall the
style to the extent needed for the subsequent specification of the
refactoring framework.

%%%%%%%%%%%%%%%%%%%%%%%%%%%%%%%%%%%%%%%%%%%%%%%%%%%%%%%%%%%%%%%%%%%%%%%%%%%%%%

\mysubsection{Generic function types}

We want to be able to write generic functions on terms over algebraic
datatypes, that is, on term types. We want these functions to be
monadic so that we can model partiality, state passing etc.\
\cite{Wadler92}. It turns out that we need two kinds of generic
functions, namely type-preserving and type-unifying ones. The former
kind of function is suitable for transforming a term while preserving
its type whereas the latter is suitable for analysing or reducing a
term. Type-preservation adheres to the type scheme $\forall x. x \to
m\ x$ where $m$ is the type-constructor parameter for the
monad. Similarly, type-unification for a given type $a$ adheres to the
scheme $\forall x. x \to m\ a$. These type schemes should not be
interpreted in the restrictive manner of parametric polymorphism.  We
need generic functions which also allow for generic traversal and
ad-hoc cases. In the present paper, we detach ourselves from the
modelling details for generic functions in Haskell. We assume that the
generic function types for \emph{t}ype-\emph{p}reserving and
\emph{t}ype-\emph{u}nifying functions are available via two Haskell
datatypes:

\[\begin{array}{llll}
\textbf{data} & \emph{Monad}\ m \Rightarrow \emph{TP}\ m & = & \ldots\\
\textbf{data} & \emph{Monad}\ m \Rightarrow \emph{TU}\ a\ m & = & \ldots
\end{array}\]

We can formulate the aforementioned intuitive type-schemes as a
contract as follows. We assume two dedicated combinators for generic
function application---one for type-preserving functions, and another
for type-unifying functions:
\[\begin{array}{lll}
\emph{applyTP} & :: & (\emph{Monad}\ m, \emph{Term}\ t) \Rightarrow \emph{TP}\ m \to t \to m\ t\\
\emph{applyTU} & :: & (\emph{Monad}\ m, \emph{Term}\ t) \Rightarrow \emph{TU}\ a\ m \to t \to m\ a          
\end{array}\]
In these type declarations, the class \emph{Term} comprises all term
types.  Let us read the declaration of \emph{applyTP}: If a
type-preserving function is applied to a term of type $t$, then the
result is also of type $t$, or of type $m\ t$ if we are precise and
account for the monadic style; similarly for type-unifying function.
Note that we need special application combinators because our generic
functions are not plain Haskell functions. They are rather opaque
terms of type $\emph{TP}\ m$ and $\emph{TU}\ a\ m$. We will gradually
rehash a few more ordinary function combinators to may use them for
generic functions, too.

%%%%%%%%%%%%%%%%%%%%%%%%%%%%%%%%%%%%%%%%%%%%%%%%%%%%%%%%%%%%%%%%%%%%%%%%%%%%%%

\mysubsection{Function combinators}

In Figure~\ref{F:StrategyPrimitives}, we provide a complete list of
all basic functions combinators we need. Let us explain these
combinators block-wise. The first block deals with combinators as they
are known from (parametric) polymorphic programming. In fact, we can
provide parametric polymorphic ``prototypes'' for the functions
combinators in the first block:

\input{literate/prototypes.math}

The prototypes embody familiar patterns in (monadic) functional
programming.  The actual details of lifting the prototypes to the
generic combinator level are omitted since we do not discuss the actual
definitions of the datatypes $\emph{TP}\ m$ and $\emph{TU}\ a\
m$. Comparing the list of prototypes and the generic combinators from
the first block, one can see that most prototypes can be instantiated
for both the type-preserving and the type-unifying case with the only
exceptions being \emph{idTP} and \emph{constTU}. This is because the
identity function is necessarily type-preserving, and the constant
function is unavoidably type-unifying.

\begin{figure}[t!]
\framebox{\parbox{.98\textwidth}{%
\vspace{20\in}%
\input{%
literate/combinators.math}\hfill
\vspace{-60\in}}}
\caption{%
Basic combinators for generic functions}
\label{%
F:StrategyPrimitives}
\end{figure}

The second block in Figure~\ref{F:StrategyPrimitives} provides
combinators for generic traversal. The \emph{all} couple applies the
argument function to \emph{all} immediate subterms (say,
children). The \emph{one} couple applies the argument function to
\emph{one} immediate subterm. The type-preserving combinators
\emph{allTP} and \emph{oneTP} preserve the outermost term constructor.
In the type-unifying case, the overall shape of the input term cannot
be preserved for simple typing arguments. As for \emph{oneTU}, one
immediate subterm is processed, and this gives immediately the result
of the type-unifying traversal. As for \emph{allTU}, all children are
processed and the intermediate results are reduced with the binary
operator of a monoid.  Hence, in this case, the unified result type
has to correspond to a monoid.

The third and the last block in the figure deals with function
update. The two combinators \emph{adhocTP} and \emph{adhocTU} enable
one to update a generic function so that it exposes type-dependent
behaviour. In other words, one can construct ad-hoc polymorphic
functions by a kind type case. This is indispensable for generic
traversals which are supposed to interact with the involved term
types. The idea is that one starts with a parametric polymorphic
function like \emph{idTP} or \emph{failTP}, and then establishes
specific behaviour for distinguished term types via generic function
update. That is, when an updated function is applied to a term, the
ad-hoc case (i.e., the second argument of \emph{adhocTP} or
\emph{adhocTU}) will be applied if applicable as for the type, that
is, if the updated term type coincides with the term type at
hand. Otherwise, we resort to the updated function (say, the generic
default, i.e., the first argument of \emph{adhocTP} or
\emph{adhocTU}).

%%%%%%%%%%%%%%%%%%%%%%%%%%%%%%%%%%%%%%%%%%%%%%%%%%%%%%%%%%%%%%%%%%%%%%%%%%%%%%

\mysubsection{Strafunski in action}

To illustrate the basic combinators, let us consider some
examples. Also, we should provide some reusable traversal schemes and
other generic functions which are frequently needed. To start with,
let us give a simple example of a combinator defined in terms of
several basic ones. The following combinator \emph{combTU} lifts a
binary operation to the generic level.

\input{literate/comb.math}

\noindent
The combinator takes a binary combinator $o$, and two type-unifying
functions $s$ and $s'$. A type-unifying function is constructed which
passes through the incoming term to both $s$ and $s'$ and combines the
intermediate results of these applications via $o$.

\begin{figure}[t!]
\framebox{\parbox{.98\textwidth}{%
\vspace{20\in}%
\input{%
literate/selectStatement1.math}\hfill
\vspace{-60\in}}}
\caption{%
A function that selects Java-like statements in a focus}
\label{%
F:selectStatement1}
\end{figure}

Let us consider a slightly more involved example related to the topic
of refactoring. In fact, we consider an example dealing with a
primitive operation involved in some Java refactorings. In
Figure~\ref{F:selectStatement1}, a function $\Varid{selectStatement}$
is defined which, given a term, looks up the statement which the focus
is placed on if any. To this end, we assume that focused statements
are surrounded by the term constructor $\Conid{StatementFocus}$. The
function $\Varid{selectStatement}$ defines a local type-unifying
function $\Varid{selectStatementStrategy}$.  The function obviously
has to be specific about statements. For that reason, we use
\emph{adhocTU} to combine a \emph{Statement}-case and a default
case. The specific case actually examines the given statement in order
to unwrap the focus term constructor if present. As for the function
constructed with \emph{adhocTU}, there are two options how failure
might arise. Either we are not faced with a statement altogether (cf.\
$\Varid{failTU}$), or the focus is not placed on the given statement
(cf.\ catch-all in \textbf{case}). The top-level application of
\emph{choiceTU} makes it possible to recover from failure. The second
alternative in the choice recursively descends into the given term via
an \emph{oneTU} traversal.

It is worth mentioning that the above problem of locating a term in a
focus can be expressed in a more generic, and hence more reusable
fashion. We will ultimately attempt such more generic definitions. In
this manner, we will approach to a style of generic refactoring.  The
present definition of \emph{selectStatement} is not generic because it
talks explicitly about statements, about the focus term constructor
\emph{StatementFocus} for statements, and it defines a traversal
scheme from scratch. By generic refactoring we mean that
language-independent refactoring functionality is identified, and that
reusable and completely generic traversal schemes are employed.

In Figure~\ref{F:StrategyLib}, we show a fragment of a library for
generic programming. These reusable application-independent
combinators are needed in the paper. Before we explain all these
definitions, let us point out a convention used in the figure. We use
names ending on ``...$S$'' for combinators which can be overloaded for
the type-preserving and the type-unifying case. This includes the
basic combinators \emph{seqS}, \emph{letS}, \emph{failS},
\emph{choiceS}, \emph{allS}, \emph{oneS}, and \emph{adhocS} introduced
separately for $\emph{TP}\ m$ and $\emph{TU}\ a\ m$ before. We can
still use all these operators with the ``..\emph{TP}'' and
``..\emph{TU}'' prefixes, if we want to express that they are used in
a context that specifically requires $\emph{TP}\ m$ or $\emph{TU}\ a\
m$.

\begin{figure}[t!]
\framebox{\parbox{.98\textwidth}{%
\vspace{20\in}%
\input{%
literate/StrategyLib.math}\hfill
\vspace{-60\in}}}
\caption{%
Specifications of some reusable generic functions}
\label{%
F:StrategyLib}
\end{figure}

Let us briefly explain the definitions in Figure~\ref{F:StrategyLib}.
The first combinator \emph{monoS} lifts a function on a term type to
the generic level by assuming failure as generic default. In fact,
failure is a prominent kind of generic default. Ultimately, we define
a few (overloaded) traversal schemes. Firstly, we define the schemes
\emph{oncetdS} and \emph{oncebuS} which attempt to apply the given
function argument once somewhere in the tree. These two schemes simply
differ in the vertical direction of search. That is, \emph{oncetdS}
and \emph{oncebuS} stand for ``once top-down'' or ``once bottom-up'',
respectively.  The schemes \emph{aboveS} is concerned with paths in
trees (say, terms). The combinator takes a generic function $s$ for
selection or transformation, and another generic function $s'$ serving
as a kind of generic predicate. The goal is to apply $s$ to a node
above another node which meets the condition $s'$. In order to
minimise the distance between the two nodes, the overall traversal is
dominated by a bottom-up traversal to find the bottom-most node
admitting application of $s$ while $s'$ is met below this node, that
is, the condition is checked in top-down manner. The last traversal
scheme \emph{propagateS} favours top-down traversal as \emph{oncetdS}
does, but in addition propagation is performed. The scheme takes an
initial parameter $e$, a type-unifying scheme $s'$ to update the
parameter before descending into the children, and the actual scheme
$s$ for selection or transformation. As an aside, the scheme
illustrates that we do not necessarily need to employ the monad
parameter for effects like propagation (or accumulation as well) but
the effect handling can be largely hidden in the traversal scheme.

As a simple exercise for applying the defined combinators, let us
rephrase the function \emph{selectStatement} from
Figure~\ref{F:selectStatement1}. We employ \emph{monoTU} to lift the
\textbf{case}\ for focus identification to the generic level. We also
use \emph{oncetdS} to describe the traversal underlying focus
selection. The resulting code is much more concise:

\input{literate/selectStatement2.math}

%%%%%%%%%%%%%%%%%%%%%%%%%%%%%%%%%%%%%%%%%%%%%%%%%%%%%%%%%%%%%%%%%%%%%%%%%%%%%%

\mysection{The refactoring framework}
\label{S:framework}

The framework for refactoring is structured as follows. Firstly, there
are several generic algorithms to perform simple analyses and
transformations as needed in the course of refactoring, e.g., to
operate in a focus, or determine free variables in a certain
scope. Secondly, there is an interface to deal with abstractions of a
language. Ultimately, we can define refactorings in terms of the
abstraction interface and the generic algorithms. The specifications
of both the generic algorithms and the refactorings carry formal
parameters which need to be instantiated to obtain
language-dependent variants. These parameters and the obligation to
provide instances of the abstraction interface form the hot spots of
the refactoring framework. For brevity, we only provide detailed
discussions of two examples of parameterized program refactorings,
namely extraction and introduction.

%%%%%%%%%%%%%%%%%%%%%%%%%%%%%%%%%%%%%%%%%%%%%%%%%%%%%%%%%%%%%%%%%%%%%%%%%%%%%%

\mysubsection{Generic algorithms}

A specification of a refactoring should preferably be composed from
simple reusable transformations. In addition, a refactoring employs
analyses to determine parameters required for some transformation
step, or to ensure some pre- or postcondition. Corresponding generic
algorithms are presented in the sequel. Firstly, we discuss
functionality to operate on a focus or a scope. Secondly, we provide
algorithms to determine free or bound names in different manners.

\begin{figure}[t!]
\framebox{\parbox{.98\textwidth}{%
\vspace{20\in}%
\input{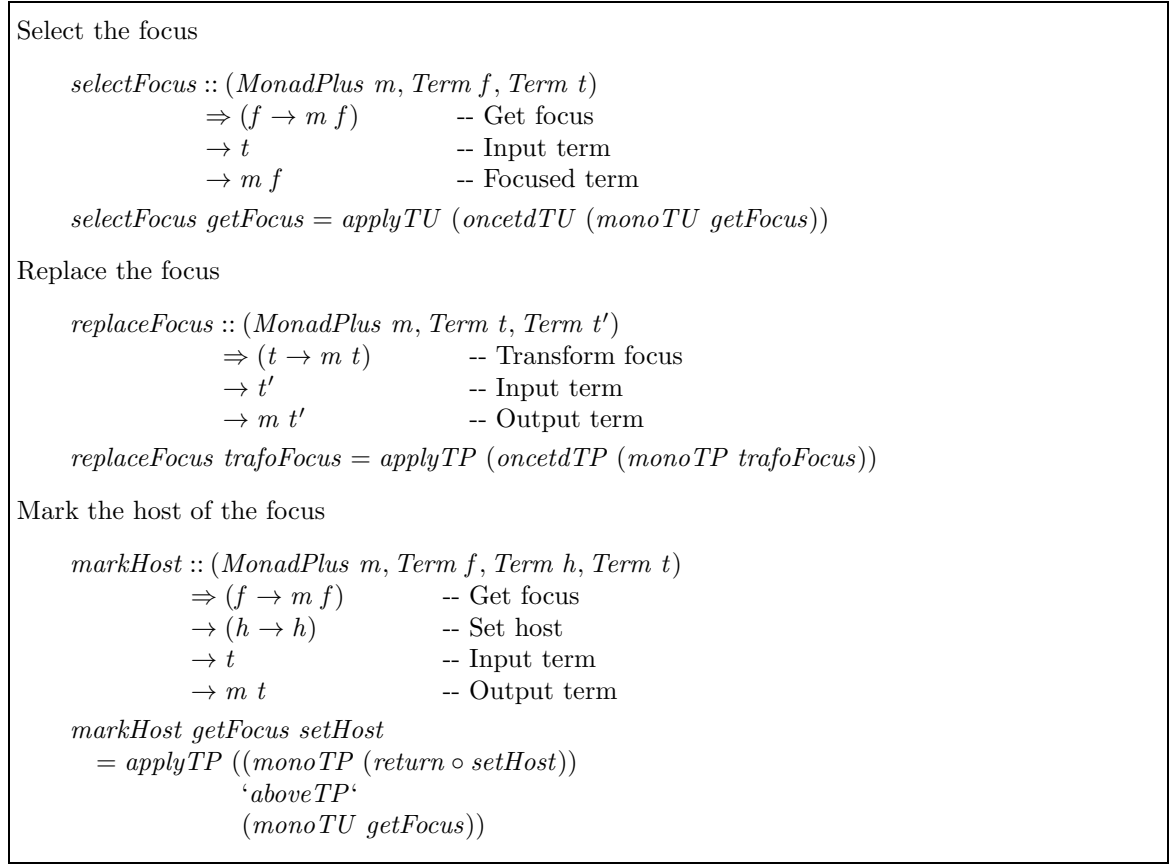}\hfill
\vspace{-60\in}}}
\caption{%
Functions to deal with focus and scope}
\label{%
F:MonoTheme}
\end{figure}

\myparagraph{Focus and scope}

In Figure~\ref{F:MonoTheme}, we specify functions to select a focused
term, to replace a term in the focus by another term, and to mark the
host of a focused term in a certain way. Note the type of these
functions.  These are ordinary polymorphic functions but they
internally employ generic functions in order to perform traversal
for the relevant selection, replacement, or marking. Let us explain
the three functions in some detail:
\begin{itemize}
\item The function \emph{selectFocus} takes a parameter
\emph{getFocus} the type of which also regulates the type of the
focused entity. The monomorphic function \emph{getFocus} is lifted to
the generic level via \emph{monoTU}. Applying the resulting generic
function to a term, it will succeed (and return the input term) if
the focus is placed on the given term. Otherwise, the application
fails. In order to apply \emph{getFocus} all over the tree until it
succeeds, we simply employ the traversal scheme \emph{oncetdS} from
Figure~\ref{F:StrategyLib}.
\item Replacement of the focused entity, as defined by the function
\emph{replaceFocus}, is very similar to selection, that is, we
basically perform a top-down traversal with one intended application
of a monomorphic function. This time, it is a type-preserving
traversal.  One might argue that the function for replacement of the
focus is unnecessarily liberal in that it further descends into terms
even if the focus was found but the replacement failed because of an
applicability condition which did not hold. Indeed, one can specify a
variant which does not descend any further once the focus has been
located. We omit this optimization.
\item The function \emph{markHost} attempts to find a term which
passes the \emph{setHost} parameter, and which is above the focused
entity identified by the \emph{getFocus} parameter. It marks then the
found host so that subsequent transformations can observe a focus on
the host. In this sense, the function is concerned with both the
notion of scope (to determine a host) and focus (as for the focused
entity and the marked host). By host we mean entities like
abstractions. To identify the host of a focused term, we employ the
scheme \emph{aboveS} from Figure~\ref{F:StrategyLib}. Here we assume
that the host of a focused term is the deepest term which meets the
following two conditions.  Firstly, it is a host-like term, that is,
it can be transformed via \emph{setHost}. Secondly, it contains the
focused term.
\end{itemize}

\myparagraph{Name analyses}

In addition to generic algorithms dealing with focus and scope, we
also need further algorithms to analyse the names used in certain ways
in program fragments. Essentially, refactorings need to be able to
determine free and bound variables in a given scope. Here we make
several assumptions. Firstly, names arise from all kinds of
abstractions available in the language at hand. Secondly, the
programming language is free to regulate name space issues, that is,
the abstractions might live in one name space, or in separated name
spaces. Thirdly, as for typed languages, abstractions can be
associated with types which are either prescribed in the input
program, or inferred by a corresponding algorithm. Fourthly, we
basically distinguish two kinds of occurrences of names, namely
declared or referring occurrences. Based on these assumptions, generic
name analyses relevant for refactoring are specified in
Figure~\ref{F:NameTheme}. As an aside, the aforementioned assumptions
are also taken into account in the interface for abstraction that
will be presented shortly. Note the types of the functions for name
analyses. These functions receive generic function parameters in order
to generically identify names and possibly their types in given terms
in a language-specific manner. Otherwise, these functions are ordinary
polymorphic functions from terms to lists (say, sets) of names (maybe
paired with types). Of course, the functions employ internally generic
functions to perform the deep collections required for name analyses.

\begin{figure}[t!]
\framebox{\parbox{.98\textwidth}{%
\vspace{20\in}%
\input{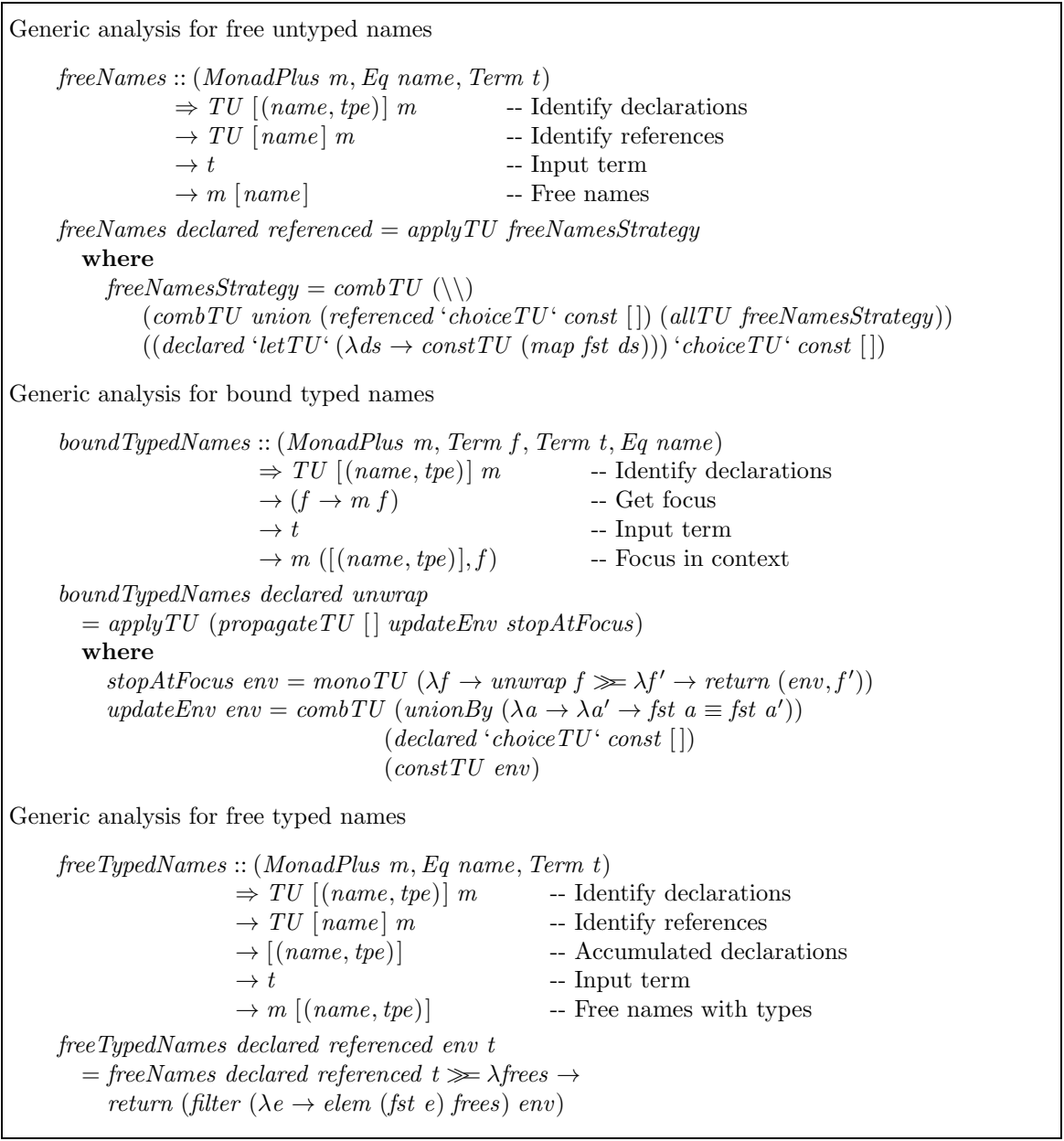}\hfill
\vspace{-60\in}}}
\caption{%
Name analyses}
\label{%
F:NameTheme}
\end{figure}

Let us explain the three functions in Figure~\ref{F:NameTheme} in
detail:
\begin{itemize}
\item The function \emph{freeNames} determines the set of free names
in a given term. To this end, the function is parameterized by two
type-unifying functions. The function \emph{declared} is meant to
identify declaration forms, and to return the corresponding declared
names if any. In the same sense, \emph{referenced} is expected to
identify referenced names. The algorithm for free name analysis is
based on a type-unifying bottom-up traversal of the following kind.
The free names correspond to the union of the names referenced at the
root node, and all the free names found for the subtrees (cf.\
$\Varid{combTU}\;\Varid{union}$ and \emph{allTU}), except the names
declared at the present node (cf.\
$\Varid{combTU}\;({\setminus}{\setminus})$).
\item The function \emph{boundTypedNames} accumulates all bound names
and their types by descending into the given term until the focus is
found.  The accumulated name-type pairs are returned together with the
focused term.  In this manner, we determine what declarations are
visible in the focused piece of code. It is interesting to notice
that, in the context of refactoring, name analyses interact with the
focus concept. The accumulation of bound names is based on the additional
assumption that a declaring occurrence of a name will usually provide
a type for the name.
\item The function \emph{freeTypedNames} is an elaboration of the
function \emph{freeNames} making use of accumulated name-type pairs
\emph{env}. In fact, a prime candidate for accumulation is the
function \emph{boundTypedNames}. That is, the function
\emph{freeTypedNames} qualifies the free names obtained by
\emph{freeNames} according to the name-type pairs received via the
argument \emph{env}. Here, we do not assume that a referring
occurrence of a name necessarily exhibits a type for the relevant
name. The types are rather obtained from the additional \emph{env}
parameter.
\end{itemize}

%%%%%%%%%%%%%%%%%%%%%%%%%%%%%%%%%%%%%%%%%%%%%%%%%%%%%%%%%%%%%%%%%%%%%%%%%%%%%%

\begin{figure}[t!]
\framebox{\parbox{.98\textwidth}{%
\vspace{20\in}%
\input{%
literate/Abstraction.math}\hfill
\vspace{-60\in}}}
\caption{%
A class of abstractions}
\label{%
F:Abstraction}
\end{figure}

\mysubsection{Abstractions}

In addition to the hot spots provided by the above generic algorithms,
we also need an interface for language abstractions to detach
ourselves from language-specific abstractions. Abstractions are so
important because most refactorings deal with declaration forms and
applications forms of abstractions. The interface is shown in
Figure~\ref{F:Abstraction}. In fact, the interface is defined as a
highly-parameterized but otherwise completely systematic (if not
trivial) Haskell class. The class members model observers and
constructors. The class parameters are essentially place holders for
syntactical domains.  There is a parameter \emph{abstr} for the domain
of abstractions itself. There are parameters \emph{name},
\emph{formal}, and \emph{body} for the constituents of abstractions.
To be precise, the domain \emph{name} does not just model names of the
particular form of abstraction at hand but it corresponds potentially
to a sum domain of all possible forms of names for a
language. Similarly, the parameter \emph{tpe} is a place holder for
all possible types (be it an attribute type, a method profile, or
others). We assume that abstractions always admit the concept of
application. Hence, there is a corresponding parameter \emph{apply}
and another parameter \emph{actual} for the arguments of an
application. The functional dependencies~\cite{Jones01} state all the
relations between the various syntactical and other domains.  The
members of \emph{Abstraction} are intended to observe all the
ingredients of both abstractions and applications. It also provides
corresponding members for construction.  Note that formal and actual
parameter lists are constructed from lists of name-type pairs.

One can say that the definition of the class \emph{Abstraction}
corresponds to the Haskell-way of defining a signature morphism. All
the class constraints on the parameters and the functional
dependencies between the parameters effectively restrict possible
instantiations.  The use of the Haskell class mechanism provides us
with two features.  Firstly, when compared to explicit parameters, we
can reduce the number of parameters in the various generic algorithms
and refactorings since the abstraction interface is global. Secondly,
note that we can easily deal with several forms of abstractions due to
overloading.

%%%%%%%%%%%%%%%%%%%%%%%%%%%%%%%%%%%%%%%%%%%%%%%%%%%%%%%%%%%%%%%%%%%%%%%%%%%%%%

\mysubsection{Refactorings}

The refactorings for extraction and introduction are defined in full
detail.  In fact, introduction, that is, insertion of a so-far unused
abstraction, is also one of the major steps of extraction. It would
be straightforward to present the dual refactorings, namely inlining
and elimination. In the conclusion of the paper we comment on further
refactorings.

\myparagraph{Generic extraction}

The parameterized transformation function that models generic
extraction is given in Figure~\ref{F:extract}. The first six
parameters are framework parameters, that is, these parameters need to
be fixed if a concrete, language-specific refactoring for extraction
is derived. The first two parameters \emph{declared} and
\emph{referenced} correspond to the ingredients of the name
analyses. The parameter \emph{find} specifies how to find the focused
fragment which is subject to extraction. The two parameters
\emph{mark} and $\mathit{find}'$ deal with marking and selecting a
focus in lists of abstractions. This second kind of focus is relevant
for the introduction step of extraction, that is, when the newly
constructed abstraction is added to the appropriate list of
abstractions. Finally, the parameter \emph{check} anticipates that
language-specific conditions need to be checked for the focused
entity. Otherwise, the final two parameters \emph{name} and
\emph{prog} just correspond to the desired name for the new
abstraction, and the input program.

\begin{figure}[t!]
\framebox{\parbox{.98\textwidth}{%
\vspace{20\in}%
\input{%
literate/extract.math}\hfill
\vspace{-60\in}}}
\caption{%
Definition of generic extraction}
\label{%
F:extract}
\end{figure}

The actual specification of the \emph{extract} refactoring is merely a
list of small analysis, destruction, construction, and transformation
steps.  Let us just read all the 11 steps in
Figure~\ref{F:extract}. First, we navigate to the focus while
accumulating the bound names (cf.\ \emph{boundTypedNames}). Then the
language-specific requirements are tested for the focused entity (cf.\
\emph{check}). Then, the abstraction is constructed in several steps
corresponding to the smaller ingredients of the abstraction. In this
course, the free names and their types are determined for the focused
piece of code. The resulting name-type pairs serve as input for the
construction of formal (and actual) parameter lists.  The actual
insertion of the constructed abstraction is defined via the separate
refactoring \emph{introduce} the application of which is preceded by
a step to mark the relevant list of abstractions (cf.\
\emph{markHost}).  Afterwards, an application is constructed in two
steps. Ultimately, the focused fragment is replaced by the application
of the new abstraction (cf.\ \emph{replaceFocus}).

\myparagraph{Generic introduction}

In Figure~\ref{F:introduce}, the generic refactoring \emph{introduce}
is specified. In order for an inserted abstraction not to interfere
with the preexisting abstractions in a program (i.e., for the sake of
semantics preservation), the name of the new abstraction should
neither be bound nor free in the scope of the target list of
abstractions. The parameters of \emph{introduce} are the ingredients
of the variable analyses, and the recognition function for the focused
list of abstractions. These are the steps which are performed by a
generic introduction. Firstly, the relevant list of abstractions is
selected from the focus. Secondly, the name of the abstraction subject
to insertion is determined. Thirdly, the free names \emph{frees} in
the relevant list of abstractions are determined. Then, also the names
\emph{defs} of all the abstractions in the local list are collected.
Afterwards, it is tested that the name of the new abstraction is
neither contained in \emph{frees} nor \emph{defs}. Ultimately, the
list of abstractions is extended with the new abstraction (cf.\
\emph{replaceFocus}).

\begin{figure}[t!]
\framebox{\parbox{.98\textwidth}{%
\vspace{20\in}%
\input{%
literate/introduce.math}\hfill
\vspace{-60\in}}}
\caption{%
Definition of generic introduction}
\label{%
F:introduce}
\end{figure}

It is important to notice that a generic refactoring is not concerned
with all the details of the static and dynamic semantics of the
involved syntactical fragments. The result of an introduction
refactoring, for example, is not necessarily well-typed because the
abstraction might fail to fit into the focused location (as for types
of free names). The only purpose of checks in refactoring are to
ensure that semantics preservation holds. This is the reason that we
check that the introduced abstraction does not override some other
visible abstraction. Simple checks for semantics preservation and
static semantics checking are two separate concerns. Of course, we
should usually perform a subsequent check to ensure that the result of
refactoring is correct regarding the static semantics. Here, we assume
that this kind of executable language semantics is available. There
are also refactorings which are completely self-checking not just for
semantics-preservation but even for static correctness. A good example
is extraction. If the instantiation of generic extraction is properly
performed, the result of a language-specific extraction will always be
statically correct.

%%%%%%%%%%%%%%%%%%%%%%%%%%%%%%%%%%%%%%%%%%%%%%%%%%%%%%%%%%%%%%%%%%%%%%%%%%%%%%

\mysection{Instantiation for \JOOS}
\label{S:joos}

We have instantiated the framework for several languages, among them a
Haskell subset, definite clause programs, XML schemata, syntax
definitions, Pascal, and the Java subset \JOOS.\footnote{\JOOS\ was
originally designed by Laurie Hendren.  The language has been used in
various courses and research projects in the last few years in various
locations.}  In the sequel, we will discuss the \JOOS\ instance in some
detail. As an aside, in \cite{LV02-PADL}, we describe an extract
method refactoring for Java (say, \JOOS) in the Strafunski style but in
a Java-specific manner, that is, without an attempt to employ a
generic and reusable specification of extraction. It is fair to say
that the non-framework approach is much less concise when compared to
the framework approach from the present paper. The \JOOS\ instance
parallels the framework. Firstly, we refine the generic algorithms of
the framework for \JOOS.  Secondly, we provide an instance of the
abstraction interface of the framework for \JOOS\ method
declarations. Ultimately, the framework refactorings are specialised
to \JOOS\ refactorings dealing with method declarations.

%%%%%%%%%%%%%%%%%%%%%%%%%%%%%%%%%%%%%%%%%%%%%%%%%%%%%%%%%%%%%%%%%%%%%%%%%%%%%%

\mysubsection{Algorithm refinement}

Firstly, we define the kinds of focus relevant for the planned
\JOOS\ refactorings. Secondly, we refine the name analyses for \JOOS.

\myparagraph{Focus}

We need two kinds of focus for the upcoming \JOOS\
refactorings. Firstly, the focus for extraction of \JOOS\ method
declarations is concerned with statements. Secondly, the focus for
insertion of \JOOS\ method declarations is of the type of lists of
method declarations. These kinds of focus are specified in
Figure~\ref{F:JoosFocus}. We assume that the syntactical domains
\emph{Statement} and \emph{MethodDeclaration} admit corresponding
constructors \emph{StatementFocus} and \emph{MethodDeclarationFocus}.
The functions for wrapping and unwrapping a focus term constructor
are then trivially defined. These function will be useful as parameters
of the generic algorithms and the refactorings (cf.\ \emph{getFocus},
\emph{setHost}, etc.).

\begin{figure}[t!]
\framebox{\parbox{.98\textwidth}{%
\vspace{20\in}%
\input{%
literate/JoosFocus.math}\hfill
\vspace{-60\in}}}
\caption{%
Kinds of focus for \JOOS}
\label{%
F:JoosFocus}
\end{figure}

\myparagraph{Name analysis}

In Figure~\ref{F:JoosName}, the domains of \JOOS\ names and types are
identified. In fact, we restrict ourselves to forms of names and types
which are relevant for the upcoming refactorings. Furthermore,
type-unifying functions for the identification of certain kinds of
names are identified. In \JOOS, we have that variables, methods, and
method parameters all live in the same name space. Hence, the type
\emph{NameJoos} for \JOOS\ names coincides with the syntactical domain
\emph{Identifier} of the \JOOS\ language (as opposed to a disjoint
union of some types of names). As for the types relevant for the
upcoming refactoring, we separate expression types and method
types. This leads to the two alternatives in the definition of
\emph{TypeJoos}.

\begin{figure}[t!]
\framebox{\parbox{.98\textwidth}{%
\vspace{20\in}%
\input{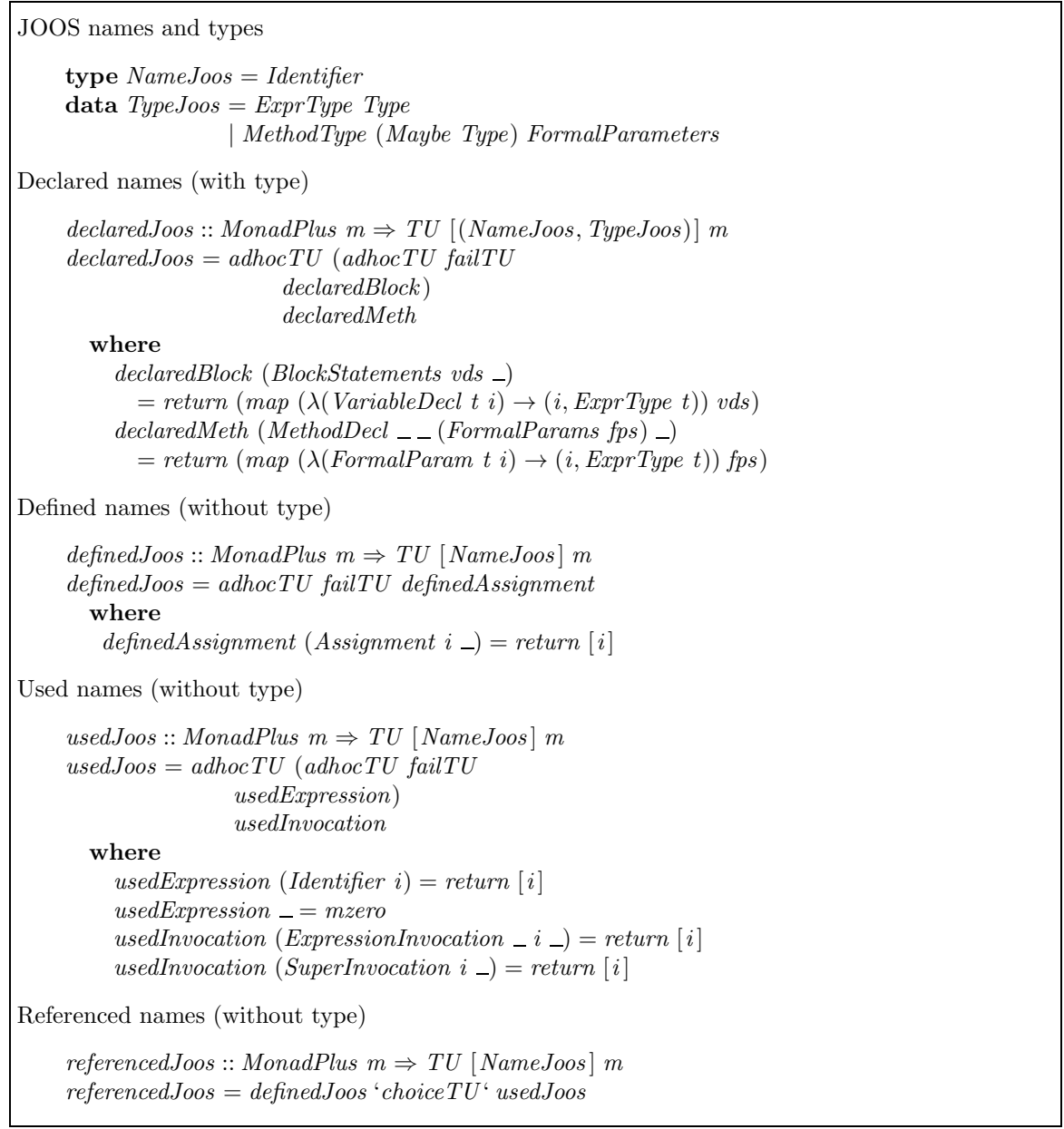}\hfill
\vspace{-60\in}}}
\caption{%
Ingredients for name analyses for \JOOS}
\label{%
F:JoosName}
\end{figure}

The framework does only separate declared and referenced variables. By
contrast, in a language like \JOOS\ where one can have side effects,
we should also separate defined (or assigned) references and using
references.  We will later see that this distinction is actually
mandatory for the correct instantiation of the \emph{extract}
refactoring. Hence, we have three generic functions for name
identification. The function \emph{declaredJoos} identifies names
together with their types. As one can see from the patterns covered by
the function, we care about variable declarations and method
parameters. The function \emph{definedJoos} identifies left-hand side
references in \JOOS\ assignments. The function \emph{usedJoos}
identifies identifiers in expressions. Again, the patterns were
selected based on a simple analysis which \JOOS\ usage patterns of
names would be relevant for the upcoming refactorings.  Finally, we
take the ``union'' of \emph{definedJoos} and \emph{usedJoos} via
\emph{choiceTU} to also be able just to identify references of any
kind (cf.\ \emph{referencedJoos}).

%%%%%%%%%%%%%%%%%%%%%%%%%%%%%%%%%%%%%%%%%%%%%%%%%%%%%%%%%%%%%%%%%%%%%%%%%%%%%%

\mysubsection{Method declarations}

In the present paper, we restrict ourselves to refactoring for \JOOS\
method declarations. The \JOOS\ language also offers other forms of
abstractions.  In particular, \JOOS\ class declarations would be
involved in many interesting refactorings. In
Figure~\ref{F:JoosAbstraction}, the framework class \emph{Abstraction}
is instantiated for \JOOS\ method declarations. The actual
specification is straightforward. Observers are more or less encoded
by pattern matching to return the corresponding fragments of a \JOOS\
method declaration; dually for the constructors. Note how the
abstraction interface and the model for the generic algorithms for
name analyses interact. Instead of plain method identifiers and types,
we use the domains \emph{NameJoos} and \emph{TypeJoos} as parameters
for the \emph{Abstraction} instance.

\begin{figure}[t!]
\framebox{\parbox{.98\textwidth}{%
\vspace{20\in}%
\input{%
literate/JoosAbstraction.math}\hfill
\vspace{-60\in}}}
\caption{%
\JOOS\ method declarations}
\label{%
F:JoosAbstraction}
\end{figure}

As an aside, in general, observers and constructors can be partial
functions (hence, the \emph{MonadPlus} constraints in
Figure~\ref{F:Abstraction}). This is useful if we want to enforce
certain side conditions on the relevant syntactical fragments.  These
side conditions can deal with normal-form issues or with other
restrictions of the framework. To give an example, consider forms of
abstractions defined by multiple equations or clauses (e.g.,
predicates in logic programming, or functions in functional
programming). If we want to determine the body of such an abstraction,
then this is only feasible (without prior normalization) if there is
precisely one equation or clause. In fact, one can think of a
refactoring to prepare abstractions accordingly, e.g., to turn a
function defined by pattern matching into a function defined in terms
of a case expression. In Figure~\ref{F:JoosAbstraction}, we use
partiality in a trivial manner, namely we require that the list of
name-type pairs only deals with expression types.

%%%%%%%%%%%%%%%%%%%%%%%%%%%%%%%%%%%%%%%%%%%%%%%%%%%%%%%%%%%%%%%%%%%%%%%%%%%%%%

\mysubsection{Refactorings refinement}

In Figure~\ref{F:JoosRefactoring}, the refactorings for extraction and
introduction of \JOOS\ method declarations are derived from the
generic ones by straight parameter passing. This is the point where
the framework approach pays off. The hot spots get closed. All the
needed ingredients for the name analyses, and for focus processing
were defined before. As for extraction, we need to define the
\JOOS-specific requirements for a valid extraction of a statement. Two
conditions need to hold (cf.\ the auxiliary function
\emph{check}). There are no return statements contained in the focused
fragment (cf.\ \emph{noReturn}). There are no free variables defined
in the focused fragment (cf.\ \emph{freeNames}).

\begin{figure}[t!]
\framebox{\parbox{.98\textwidth}{%
\vspace{20\in}%
\input{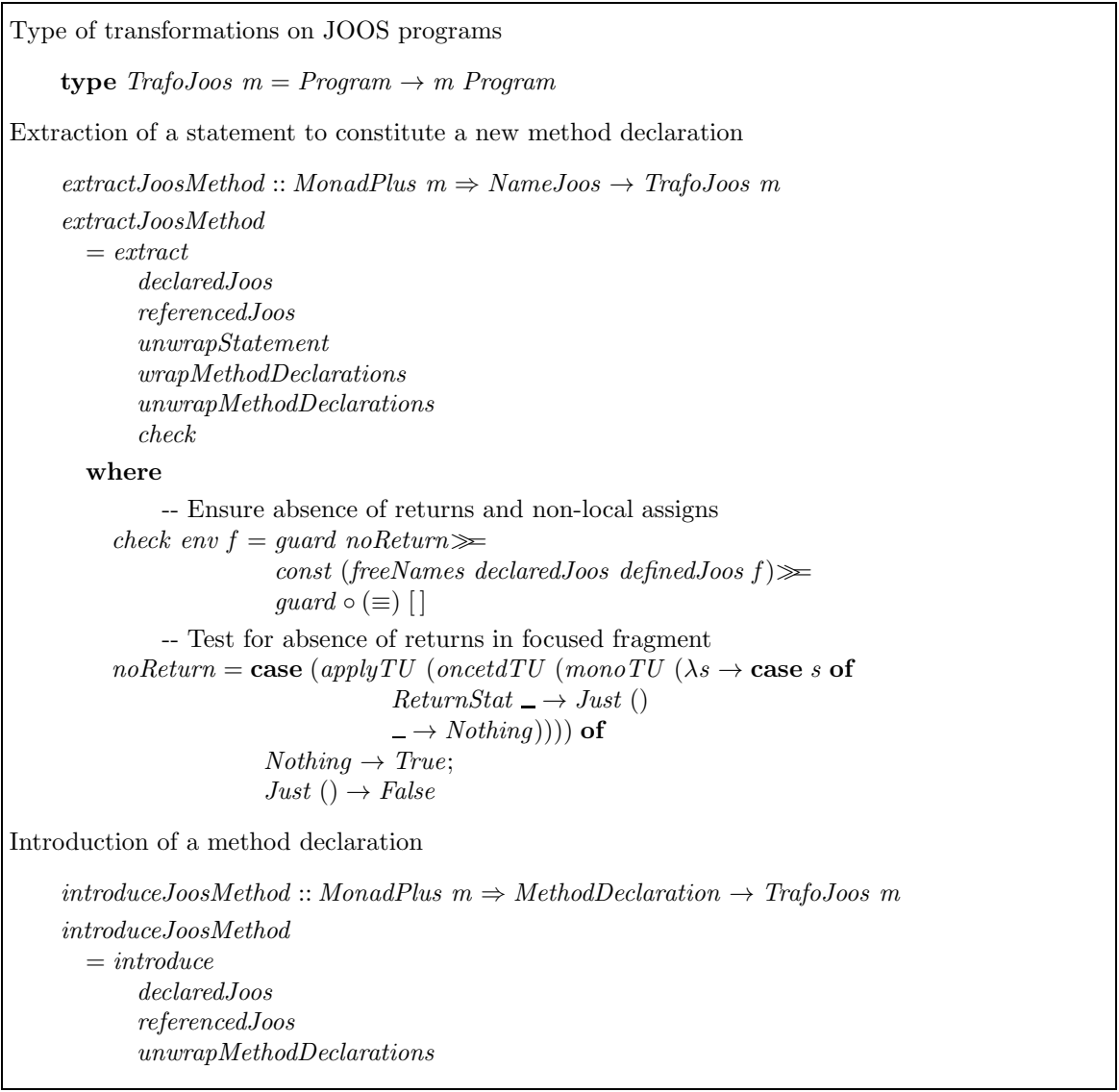}\hfill
\vspace{-60\in}}}
\caption{%
\JOOS\ extraction and introduction by specialisation}
\label{%
F:JoosRefactoring}
\end{figure}

%%%%%%%%%%%%%%%%%%%%%%%%%%%%%%%%%%%%%%%%%%%%%%%%%%%%%%%%%%%%%%%%%%%%%%%%%%%%%%

\mysection{Concluding remarks}
\label{S:concl}

\myparagraph{Contribution}

We have shown how program transformations for refactoring can be
represented in a largely language-independent manner using generic
functional programming as a sufficiently expressive and concise
specification medium. From the examples given, it is clear that
several refactorings for different forms of abstractions are
accessible for such a generic approach. Among these refactorings there
are extraction, introduction, inlining, and elimination.  The ability
to specify program transformations at this level of abstraction allows
us to capture commonalities of different programming languages in a
way which provides new insight into language design and language
semantics. One is used to the idea that frameworks for static and
dynamic semantics are meant to cover common building blocks of
languages~\cite{Moggi91,Mosses92}.  In the context of executable
language definition or language implementation~\cite{FMY92,KW94}, the
idea of reusable components is also quite common. The contribution of
the present paper is that we instantiated the idea of common building
blocks for program transformation on the basis of program
refactorings. We do not argue that a generic refactoring framework
like the one we have proposed is particularly strong because it would
enable reuse of program transformations. This would be like saying
that modular semantics has significantly simplified compiler
implementation.  It is more important that one is able to talk about
commonalities in mathematical and transformational semantics to
witness the structure underlying different programming languages.

\myparagraph{Related work}

The idea of operator suites for refactoring is, of course, not new. In
his seminal thesis \cite{OpdykePhD} and accompanying conference
papers, Opdyke develops a set of operators for refactoring
object-oriented frameworks. His results are somewhat independent of
the actual object-oriented programming language. Based on such
operator suites, corresponding refactoring tools have been
designed~\cite{Moore96,RBJ97}. As for object-oriented programming,
tool-supported refactoring is well established.  What is new in our
work is that we collect refactorings in a truly language-independent,
declarative, prescriptive and executable framework.

Research on program transformation usually aims at some degree of
language independence. In \cite{PP96}, for example, rules and
strategies for transforming both logic and functional programs are
examined in depth making only few assumptions about the covered
languages. Our transformation operators collected in the framework are
original in that the technicalities of refactoring such as focus, name
analyses, construction and destruction, or scope are all treated in a
generic manner.

An important initial contribution to the idea of generic
transformations originated from the Stratego project~\cite{VBT98}
where traversal schemes for analysis and transformation have been
identified as reusable building blocks of program transformations. In
\cite{Visser00}, basic traversal schemes but also algorithms for
variable analysis, unification, and substitution were specified in
Stratego---a language with prime support for term traversal, but
without strong typing, and support for general higher-order
functions. Our work clearly illustrates that higher-orderness and
types are desirable if not indispensable for transformation
frameworks. Higher-orderness is implied by the nature of the involved
parameters, and by the employment of higher-order functional
programming techniques for a reasonably concise style. Types are not
just convenient for documentation purposes, but they are actually
instrumented to guide traversals (recall generic function update via
\emph{adhocTP} and \emph{adhocTU}). Types are also essential to
constrain valid instantiations of the framework. Without strong type
checking, very generic, highly-parameterized frameworks are easily
configured in an inconsistent manner.

Our framework provides a model for manual, local transformations
(i.e., refactorings) aiming at some kind of improvement of the
refactored program in structural terms. Refactoring is different from
other forms of transformational programming where one is rather
interested in the \emph{calculation} of a usually efficient program
from a specification~\cite{Partsch90,BM97}.

\myparagraph{Perspective}

Besides extraction, introduction, inlining, and elimination, further
generic refactorings are conceivable, e.g., refactorings for lifting
and dropping abstractions, say to move around abstractions in nested
levels of abstraction~\cite{DS97,DS00}. Language-specific refactoring
catalogs as in \cite{OpdykePhD,Fowler99} should also be investigated
to systematically extract all refactorings which make sense at an
language-independent, abstract level.  One might also want to go beyond
refactorings in the sense that more powerful adaptations are enabled,
e.g., the adaptations from \cite{PS90,Laemmel99-PEPM} to add
computational behaviour. Furthermore, the specification of compound
(generic) transformation schemes (say, strategies) in the sense of
\cite{PP96} is a subject for future work.  Moreover, the integration
of generic refactorings and truly language-specific refactorings
deserves some effort, e.g., the very object-oriented refactorings in
\cite{Fowler99}, or the specifically functional refactorings in
\cite{Laemmel00-SFP99}, e.g., monad introduction.

\bibliographystyle{abbrv}
\bibliography{report}

\tableofcontents

\end{document}